\documentclass[9pt,twocolumn,twoside]{opticajnl}
\journal{opticajournal} 

\setboolean{shortarticle}{true}

\usepackage{lineno}

\title{Modulation-Induced Memory Resonances in Dynamically Driven Optical Cavities}

\author{ FAIZAN H. LONE}
\affil{Department of Physics, Indian Institute of Technology (IIT), Ropar, Rupanagar, Punjab 140001, India.}
\affil{faizan.21phz0022@iitrpr.ac.in}

\begin{abstract}
Temporal modulation of optical cavities provides a powerful mechanism for controlling photonic states beyond the static cavity paradigm. Here, we investigate the dynamics of periodically driven optical cavities using a time-dependent cavity-field model with modulated resonance frequencies. We demonstrate modulation-induced tuning of cavity resonances and analyze the resulting intracavity field evolution in both amplitude and phase. The interplay between periodic modulation and cavity dissipation gives rise to pronounced hysteretic behavior, revealing memory effects associated with the finite photon lifetime of the cavity. By quantifying the hysteresis area as a function of modulation frequency, we identify distinct memory resonances characterized by enhanced field-history dependence at specific driving frequencies. Furthermore, we examine the time-resolved local density of optical states (LDOS) and show that temporal modulation enables dynamic spectral redistribution of photonic states, with the degree of confinement governed by the cavity loss rate. Our results establish a simple framework for understanding and engineering memory effects in dynamically driven optical resonators and provide new opportunities for tunable photonic environments, dynamic LDOS control, and reconfigurable light–matter interactions.

\end{abstract}

\setboolean{displaycopyright}{false} 

\begin{document}

\maketitle


\section{Introduction}
Optical cavities constitute one of the most versatile platforms for manipulating electromagnetic fields and enhancing light--matter interactions \cite{vahala2003optical,peng2017enhancing,engel2012light}. Their ability to confine photons within small mode volumes while maintaining high quality factors has enabled a broad range of applications including cavity quantum electrodynamics, optical frequency-comb generation, nonlinear photonics, precision metrology, and integrated quantum technologies \cite{anderson2018highly,pasquazi2018micro,yadav2025vacuum,miller2015tunable,djorwe2024parametrically}. The continuing development of micro- and nanophotonic resonators has further stimulated interest in dynamically controllable photonic environments capable of manipulating optical states in both space and time\cite{kippenberg2007cavity,aspelmeyer2014cavity}.

Recent years have witnessed significant advances in dynamically driven and parametrically modulated optical cavities \cite{bothner2020cavity}. Temporal modulation of cavity parameters has been exploited for resonance tuning, frequency conversion, nonreciprocal photonics, Floquet engineering, and the generation of novel dissipative states of light \cite{englebert2021temporal,leonhardt2022parametrically,moille2024parametrically}. In particular, parametrically driven Kerr resonators have recently emerged as a powerful platform for realizing phase-bistable cavity solitons and pure-Kerr temporal solitons in integrated microcavities, demonstrating the rich nonequilibrium physics accessible through time-dependent cavity driving \cite{englebert2021temporal}. Such developments have established dynamic cavity modulation as a promising route toward programmable photonic functionalities \cite{goldstein2020spontaneous,deotare2012all,tian2022all}.
Alongside these advances, memory and history-dependent effects have attracted increasing attention in nonlinear optical systems. Recent work has shown that optical cavities possessing finite response times can exhibit memory-induced excitability, hysteresis, and complex dynamical transitions when the memory timescale becomes comparable to the cavity relaxation time \cite{braeckeveldt2024memory,peters2026timekeeping}. Despite these developments, the emergence of modulation-induced memory resonances in periodically driven optical cavities remains largely unexplored. In particular, the interplay between cavity dissipation, modulation frequency, and photon lifetime can generate nontrivial hysteretic dynamics and enhanced field-history dependence that are absent in static resonators.

In this Letter, we investigate the dynamics of periodically modulated optical cavities using a time-dependent cavity-field model with temporally varying resonance frequencies. We demonstrate modulation-induced tuning of cavity resonances, analyze the amplitude and phase evolution of the intracavity field, and reveal pronounced hysteresis arising from the finite response time of the cavity. By quantifying the hysteresis area as a function of modulation frequency, we identify distinct memory resonances characterized by enhanced field-history dependence at specific driving frequencies. Furthermore, we examine the time-resolved local density of optical states (LDOS) and show how temporal modulation enables dynamic control of the photonic environment. These results establish modulation-induced memory resonances as a fundamental dynamical feature of driven optical cavities and provide a route toward engineered photonic memory and tunable light--matter interactions.

To describe the cavity dynamics, we employ a time-dependent coupled-mode formalism \cite{fan2003temporal}in which the intracavity field amplitude $a(t)$ evolves according to

\begin{equation}
\frac{da(t)}{dt}
=
-\left(
\frac{\kappa}{2}
+
i\omega_c(t)
\right)a(t)
+
F ,
\label{eq:cavity}
\end{equation}

where $\kappa$ is the cavity decay rate, $F$ denotes the external driving field, and $\omega_c(t)$ represents the instantaneous cavity resonance frequency. The first term accounts for cavity losses and phase accumulation, while the second term describes continuous excitation by the external field.

The cavity resonance is assumed to be periodically modulated according to

\begin{equation}
\omega_c(t)
=
\omega_0
+
\Delta\omega
\sin(\Omega_m t),
\label{eq:modulation}
\end{equation}

where $\omega_0$ is the unperturbed cavity resonance frequency, $\Delta\omega$ is the modulation amplitude, and $\Omega_m$ is the modulation frequency. This temporal modulation introduces a dynamically varying photonic environment that continuously alters the cavity resonance condition.

The observable quantity investigated throughout this work is the intracavity intensity

\begin{equation}
I(t)
=
|a(t)|^2 ,
\label{eq:intensity}
\end{equation}

which provides direct information about the energy stored within the cavity. To characterize memory effects, we analyze the hysteretic response of the cavity by plotting $I(t)$ as a function of the instantaneous resonance frequency $\omega_c(t)$ and calculate the corresponding hysteresis area.
In addition, the time-resolved local density of optical states (LDOS) \cite{liang2013formulation}is modeled using a Lorentzian spectral response,

\begin{equation}
\rho(\omega,t)
=
\frac{2Q/\pi}
{
1
+
4Q^2
\left[
\omega-\omega_c(t)
\right]^2
},
\label{eq:ldos}
\end{equation}
where $Q=\omega_0/\kappa$ is the cavity quality factor. Equations~(\ref{eq:cavity})--(\ref{eq:ldos}) provide a simple framework for exploring the interplay between temporal modulation, cavity dissipation, memory effects, and dynamic LDOS engineering in driven optical resonators.
  
\begin{figure}
    \centering
    \includegraphics[width=1\linewidth]{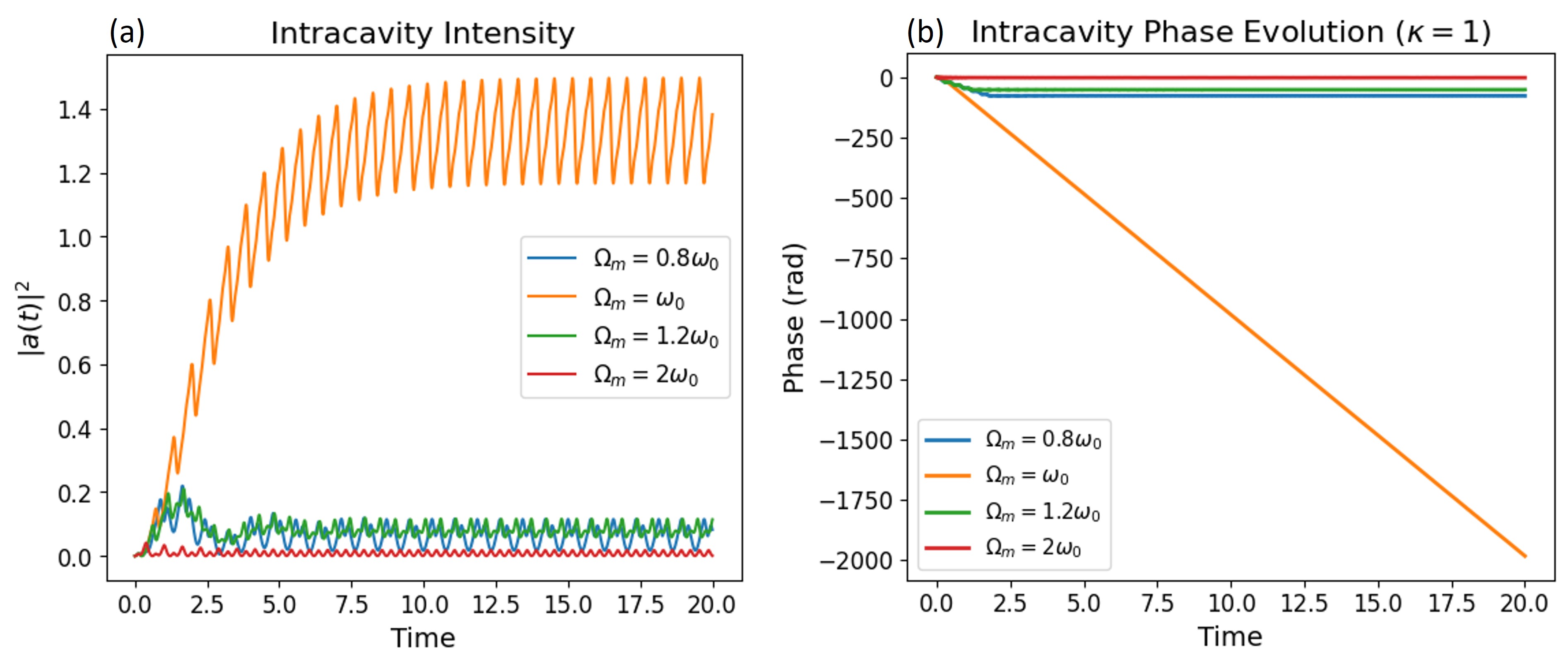}
    
    \caption{{
(a) Temporal evolution of the normalized intracavity intensity $|a(t)|^{2}$ for different modulation frequencies $\Omega_{m}$ in a dynamically modulated optical cavity with fixed cavity loss rate $\kappa=1$. A pronounced enhancement of the intracavity field is observed when the modulation frequency approaches the cavity resonance frequency ($\Omega_{m}=\omega_{0}$), indicating resonant energy transfer from the modulation to the cavity mode. Away from resonance ($\Omega_{m}=0.8\omega_{0},\,1.2\omega_{0},\,2\omega_{0}$), the cavity response remains significantly weaker. (b) Corresponding intracavity phase evolution $\phi(t)=\arg[a(t)]$. Near resonance, the phase exhibits rapid accumulation and strong temporal variation, whereas off-resonant modulation results in comparatively weak phase dynamics. These results demonstrate the simultaneous amplitude and phase sensitivity of the cavity field to periodic temporal modulation.
}}
    \label{fig:fig1}
\end{figure}
\section{Results and Discussion}

We first examine the dynamical response of the periodically modulated cavity by varying the modulation frequency $\Omega_m$ while keeping the cavity loss rate fixed at $\kappa=1$. Figure~\ref{fig:fig1} summarizes the temporal evolution of the intracavity field intensity and phase for several modulation frequencies relative to the cavity resonance frequency $\omega_0$.
A clear transition in the cavity response is observed as the modulation frequency approaches resonance. When $\Omega_m \approx \omega_0$, the intracavity field undergoes a substantial amplification accompanied by sustained oscillatory dynamics as shown in Fig.~\ref{fig:fig1}(a) . This behavior indicates efficient energy exchange between the external modulation and the cavity mode, resulting in enhanced photon accumulation inside the resonator. In contrast, modulation frequencies detuned from resonance produce significantly weaker cavity excitation, reflecting the reduced overlap between the modulation timescale and the intrinsic cavity dynamics.

\begin{figure}
    \centering
    \includegraphics[width=1\linewidth]{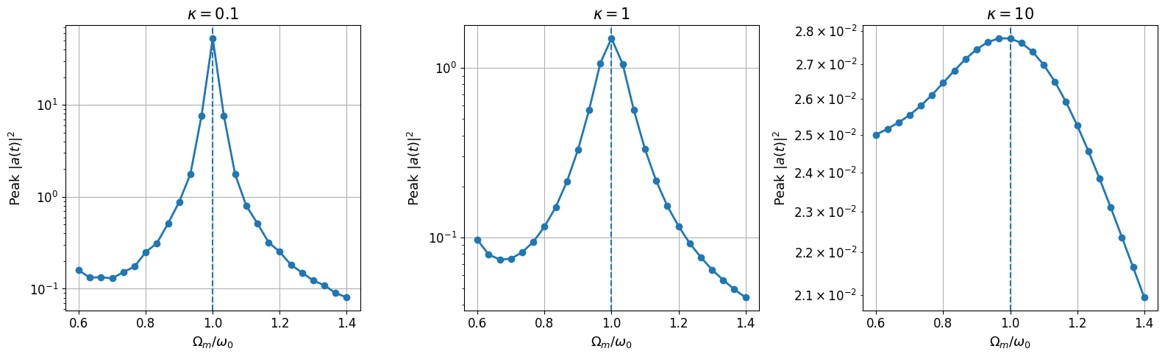}
    \caption{{
Peak intracavity intensity as a function of the normalized modulation frequency $\Omega_{m}/\omega_{0}$ for three cavity loss rates: (left) $\kappa=0.1$, (center) $\kappa=1$, and (right) $\kappa=10$. The dashed vertical line indicates the resonance condition $\Omega_{m}=\omega_{0}$. For low-loss cavities ($\kappa=0.1$), a sharp resonance peak is observed, corresponding to efficient energy transfer from the temporal modulation to the cavity mode. As the cavity loss increases, the resonance progressively broadens and its amplitude decreases, reflecting the reduced photon lifetime and diminished field buildup within the cavity. In the strongly damped regime ($\kappa=10$), the resonance response becomes weak and spectrally broadened, indicating suppression of modulation-induced enhancement. These results demonstrate the critical role of cavity dissipation in governing the frequency-selective response of dynamically modulated optical resonators.
}}
     \label{fig:fig2}
\end{figure}

The phase evolution provides complementary information regarding the dynamical state of the cavity as shown in Fig.~\ref{fig:fig1}(b) . Near resonance, the intracavity field exhibits strong phase accumulation over time, signaling a substantial modification of the optical path experienced by the circulating field. Away from resonance, the phase remains comparatively stable, indicating that the cavity follows the modulation only weakly. The simultaneous enhancement of both amplitude and phase responses demonstrates that the modulation frequency acts as a key control parameter governing the cavity dynamics.These observations establish the existence of a modulation-induced resonance in the driven cavity and motivate a more systematic investigation of its frequency dependence. To quantify this behavior, we next examine the peak intracavity intensity over a continuous range of modulation frequencies and cavity loss rates. To quantify the modulation-induced resonance identified in Fig.~\ref{fig:fig1}, we evaluate the peak intracavity intensity as a function of the normalized modulation frequency $\Omega_m/\omega_0$ for different cavity loss rates. The resulting resonance spectra are shown in Fig.~\ref{fig:fig2}.
The cavity exhibits a pronounced frequency-selective response, with the maximum intracavity intensity occurring close to the resonance condition $\Omega_m=\omega_0$. This confirms that the strongest energy transfer takes place when the external modulation timescale matches the intrinsic cavity frequency. The resonance behavior is strongly influenced by the cavity dissipation rate. In the weak-loss regime, photons remain stored within the cavity for many modulation cycles, allowing constructive accumulation of energy and producing a highly localized resonance response. As the loss rate increases, the reduced photon lifetime limits the ability of the cavity field to build up coherently, leading to a substantial suppression of the resonant enhancement.
The observed evolution from a sharp resonance to a broad and weak response reflects the competition between temporal driving and dissipative relaxation. This behavior is analogous to the reduction of the quality factor in conventional resonators, where increasing losses decrease both the spectral selectivity and energy-storage capability of the system. Consequently, cavity dissipation emerges as a key parameter controlling the effectiveness of modulation-induced energy localization.

The strong dependence of the resonance characteristics on $\kappa$ suggests that the cavity response cannot be described solely by the instantaneous resonance condition. Instead, the finite cavity response time introduces a history dependence in the field dynamics. To explore this effect, we next investigate the evolution of the cavity state in the $(\omega_c,|a(t)|^2)$ phase space and analyze the emergence of hysteresis and memory effects under periodic modulation.
The dissipation-dependent resonance behavior discussed above suggests that the cavity response is governed not only by the instantaneous value of the modulation but also by the finite time required for the intracavity field to adapt to changes in the cavity resonance. To investigate this temporal lag, we examine the cavity dynamics in the phase space defined by the instantaneous resonance frequency $\omega_c(t)$ and the intracavity intensity $|a(t)|^2$.

\begin{figure}
    \centering
    \includegraphics[width=1\linewidth]{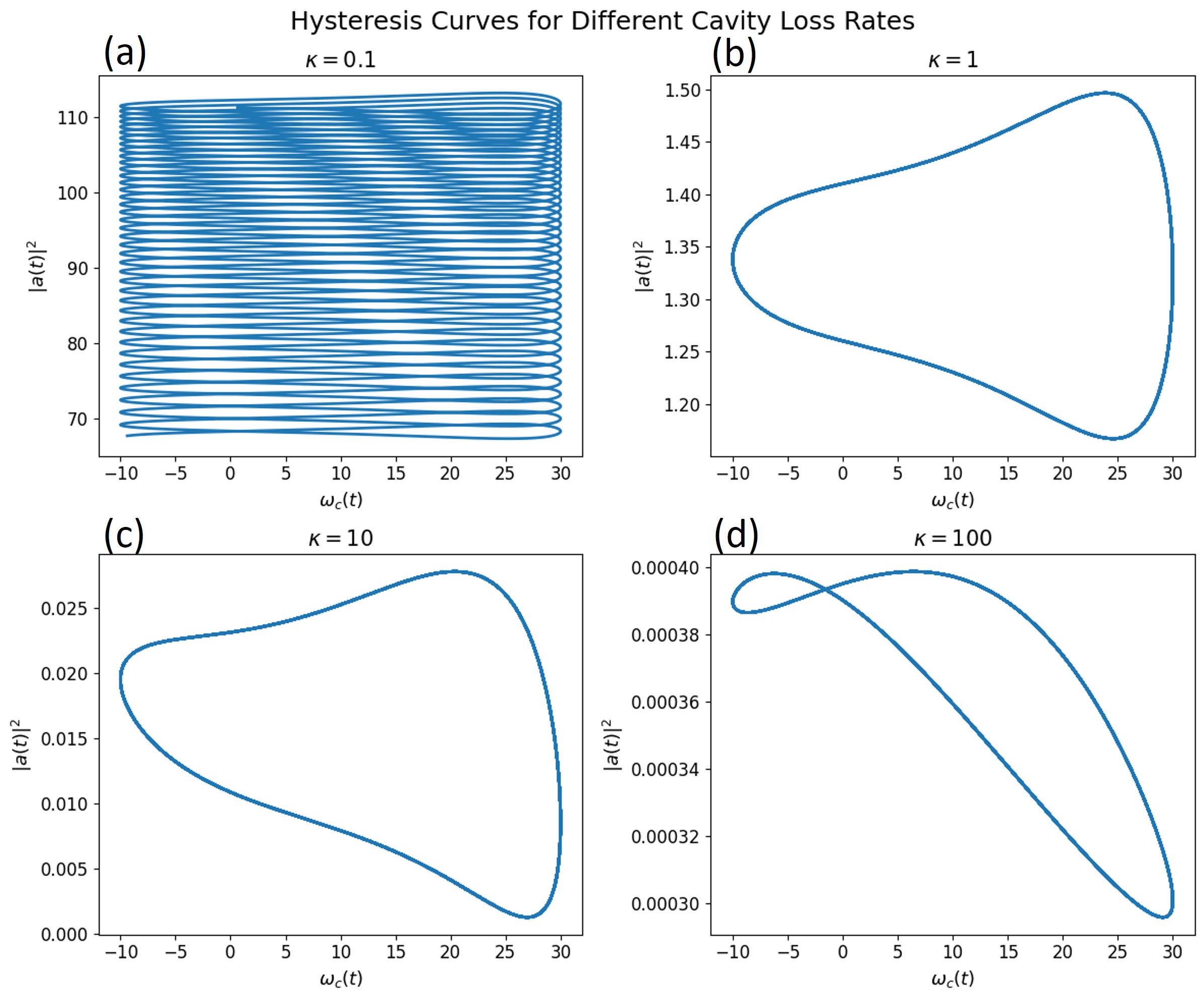}
    \caption{{
Hysteresis loops obtained by plotting the intracavity intensity $|a(t)|^{2}$ as a function of the instantaneous cavity resonance frequency $\omega_{c}(t)$ for different cavity loss rates: (a) $\kappa=0.1$, (b) $\kappa=1$, (c) $\kappa=10$, and (d) $\kappa=100$. The closed-loop trajectories arise from the finite response time of the intracavity field to the periodic modulation of the cavity resonance, indicating the presence of dynamical memory effects. For weak dissipation ($\kappa=0.1$), the cavity exhibits large-amplitude oscillations and pronounced hysteresis due to the long photon lifetime and strong field accumulation. As the cavity loss increases, the loop area progressively decreases, reflecting the reduced ability of the cavity to retain information about its previous states. In the strongly damped regime ($\kappa=100$), only a small residual hysteresis remains, indicating that the cavity field rapidly follows the external modulation with minimal temporal lag. These results demonstrate the transition from a memory-dominated regime in high-$Q$ cavities to a nearly adiabatic response in low-$Q$ cavities.
}} 
 \label{fig:fig3}
\end{figure}

Figure~\ref{fig:fig3} shows the resulting trajectories for several cavity loss rates. Rather than retracing the same path during successive modulation cycles, the system forms closed hysteresis loops. This behavior demonstrates that the cavity state depends on its dynamical history, providing direct evidence of memory effects induced by the interplay between periodic driving and finite photon lifetime.
The size and shape of the hysteresis loops depend strongly on the cavity dissipation rate. In weakly dissipative cavities, the long photon lifetime allows the intracavity field to retain information over many modulation periods, producing large deviations between the forward and backward evolution of the cavity state. As the loss rate increases, the cavity relaxes more rapidly toward its instantaneous steady state, reducing the temporal lag between the modulation and the field response. Consequently, the hysteresis loops become progressively smaller, indicating a gradual loss of memory.

These observations reveal a continuous transition between two dynamical regimes. At low dissipation, the cavity operates in a memory-dominated regime where the field evolution is strongly influenced by its past history. In contrast, highly dissipative cavities exhibit a nearly adiabatic response in which the field follows the external modulation with only a small residual delay. The hysteresis loop area therefore provides a natural quantitative measure of the strength of the memory effect.
To characterize this behavior more systematically, we next calculate the hysteresis area as a function of modulation frequency and cavity loss rate, enabling the identification of memory resonances associated with enhanced history dependence in the driven cavity dynamics.
Having established the existence of hysteresis and memory effects in the driven cavity dynamics, we now quantify the strength of this memory by evaluating the area enclosed by the hysteresis loops. The hysteresis area provides a direct measure of the temporal lag between the cavity response and the external modulation and therefore serves as an effective metric for dynamical memory.

\begin{figure}
    \centering
    \includegraphics[width=1\linewidth]{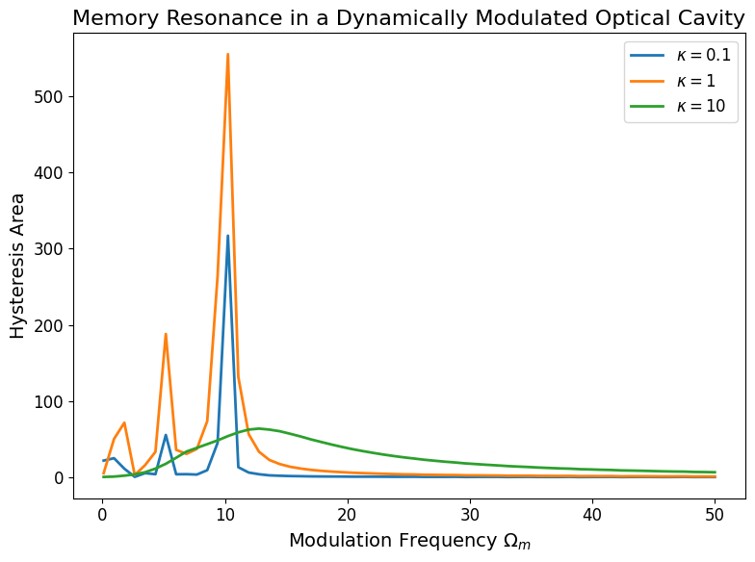}
    \caption{{
Memory resonance in a dynamically modulated optical cavity quantified through the hysteresis area as a function of the modulation frequency $\Omega_{m}$ for different cavity loss rates $\kappa$. The hysteresis area serves as a measure of the cavity's dynamical memory and temporal lag in response to periodic resonance modulation. Distinct resonance peaks emerge at specific modulation frequencies where the interplay between the modulation timescale and the photon lifetime maximizes the memory effect. For weak and moderate dissipation ($\kappa=0.1$ and $\kappa=1$), sharp resonant enhancements of the hysteresis area are observed, indicating strong non-adiabatic cavity dynamics and significant field-history dependence. As the cavity loss increases ($\kappa=10$), the resonance broadens and its amplitude is substantially reduced, reflecting the diminished ability of the cavity to store electromagnetic energy and retain memory of previous states. These results reveal the existence of modulation-induced memory resonances and demonstrate that both the strength and spectral width of the memory response can be engineered through cavity dissipation.
}}
 \label{fig:fig4}
   
\end{figure}

Figure~\ref{fig:fig4} presents the hysteresis area as a function of the modulation frequency for several cavity loss rates. Rather than varying monotonically with driving frequency, the memory measure exhibits pronounced resonant features, indicating that memory effects are selectively enhanced at particular modulation frequencies. This behavior reveals the existence of \emph{memory resonances}, where the modulation timescale becomes optimally matched to the intrinsic cavity relaxation dynamics.
The emergence of these resonances can be understood from the competition between periodic driving and photon decay. When the modulation is too slow, the cavity field can adiabatically follow the instantaneous resonance frequency, resulting in only weak hysteresis. Conversely, for very rapid modulation the cavity cannot efficiently accumulate energy before the resonance condition changes, again leading to a reduced memory response. Maximum hysteresis therefore occurs in an intermediate regime where the modulation period becomes comparable to the photon lifetime, producing the strongest history dependence.

The dependence of the resonance profile on the cavity loss rate further highlights the role of dissipation in controlling memory formation. Cavities with lower loss rates exhibit stronger and sharper memory resonances due to their longer photon storage times, whereas increased dissipation suppresses the buildup of memory and broadens the resonance response. These results establish cavity dissipation as a key parameter for engineering memory effects in dynamically driven photonic systems.The existence of modulation-induced memory resonances suggests that dynamically modulated cavities can operate as tunable photonic memory elements whose response can be optimized through the interplay of modulation frequency and photon lifetime. To gain further insight into the underlying photonic-state dynamics associated with these memory resonances, we next examine the time-dependent evolution of the local density of optical states.

Having established that modulation-induced memory effects are governed by the interplay between the modulation timescale and cavity dissipation, we finally examine how these dynamics influence the photonic environment itself through the local density of optical states (LDOS). Since the LDOS determines the spectral availability of photonic modes and directly controls spontaneous emission rates and light--matter coupling strengths, its temporal evolution provides important insight into the consequences of dynamic cavity modulation.

Figure~\ref{fig:5} presents the time-resolved LDOS for three representative cavity loss rates. The instantaneous cavity resonance frequency follows the periodic trajectory defined by Eq.~(\ref{eq:modulation}), shown by the dashed curve. The LDOS maximum continuously tracks this moving resonance, demonstrating that temporal modulation dynamically reshapes the spectral distribution of available photonic states.

A pronounced dependence on the cavity loss rate is observed. For weak dissipation ($\kappa=0.1$), the LDOS remains strongly localized around the resonance trajectory, indicating narrow spectral linewidths and efficient photon confinement. In this regime, the cavity supports highly selective optical modes whose frequencies can be dynamically tuned while preserving strong spectral localization. Increasing the cavity loss rate to $\kappa=1$ produces a noticeable broadening of the LDOS distribution, reflecting the reduced photon lifetime and the corresponding decrease in spectral selectivity.
The effect becomes particularly evident in the strongly damped regime ($\kappa=10$), where the LDOS extends over a broad frequency range throughout the modulation cycle. Rather than supporting sharply defined resonances, the cavity provides a spectrally diffuse photonic environment whose modal structure evolves continuously in time. The broadening originates from the increased cavity linewidth associated with stronger dissipation and illustrates how cavity losses directly influence the temporal redistribution of photonic states.
\begin{figure}
    \centering
    \includegraphics[width=1\linewidth]{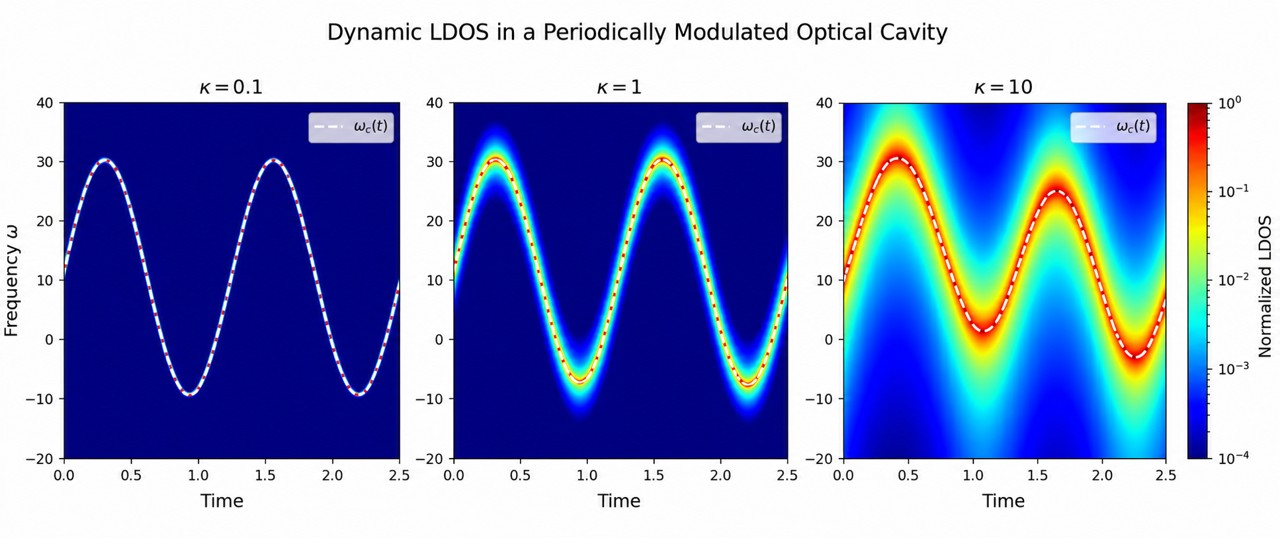}
    \caption{{
Time-resolved local density of states (LDOS) of a periodically modulated optical cavity for different cavity loss rates: (left) $\kappa=0.1$, (center) $\kappa=1$, and (right) $\kappa=10$. The color map represents the normalized LDOS, while the dashed white curve denotes the instantaneous cavity resonance frequency $\omega_{c}(t)$. For low cavity loss ($\kappa=0.1$), the LDOS is strongly localized around the resonance trajectory, reflecting a high-$Q$ cavity with excellent spectral selectivity and strong photon confinement. As the cavity loss increases, the LDOS progressively broadens in frequency, indicating reduced photon lifetime and enhanced spectral overlap. In the strongly damped regime ($\kappa=10$), the resonance becomes substantially broadened, producing a dynamically tunable but spectrally diffuse photonic environment. These results demonstrate how temporal modulation and cavity dissipation jointly govern the spatiotemporal evolution of the photonic density of states, enabling dynamic control of light--matter interactions and emission processes in time-dependent optical resonators.
}}
    \label{fig:fig5}
\end{figure}

These results reveal that temporal modulation and cavity dissipation jointly determine the dynamical structure of the photonic density of states. While high-$Q$ cavities enable sharply localized and dynamically tunable resonances, increased losses progressively transform the LDOS into a broader and less selective spectral distribution. Such dynamic control of the LDOS offers opportunities for tailoring spontaneous emission processes, manipulating Purcell enhancement, and engineering time-dependent light--matter interactions in reconfigurable photonic systems.

\section{Conclusion}

In conclusion, we have investigated the dynamics of periodically modulated optical cavities and demonstrated the emergence of modulation-induced memory resonances arising from the interplay between temporal driving and cavity dissipation. We showed that resonant modulation enhances intracavity field buildup and produces pronounced hysteretic behavior associated with the finite photon lifetime of the cavity. By quantifying the hysteresis area, we identified distinct memory resonances where the cavity exhibits maximum history-dependent response. Furthermore, we demonstrated that cavity dissipation governs both the strength of the memory effect and the temporal evolution of the local density of optical states, enabling dynamic control of the photonic environment. These findings establish dynamically driven optical cavities as a simple yet versatile platform for engineering photonic memory, tunable LDOS landscapes, and adaptive light--matter interactions, with potential applications in cavity quantum electrodynamics, dynamic Purcell enhancement, and reconfigurable nanophotonic devices.

\noindent{\large\textbf{Data availability.}} Data underlying the results presented in this paper are not publicly available at this time but may be obtained from the authors upon reasonable request.\\

\bibliography{bib}
\end{document}